\documentstyle[11pt,newpasp,twoside,psfig]{article}
\begin{document}

\title{Efficiency of stripping mechanisms}
\author{Fran\c{c}oise Combes}
\affil{Observatoire de Paris, 61 Av. de l'Observatoire,
F-75 014, Paris, France}

\setcounter{page}{411}
\index{Combes, F.}

\begin{abstract}
There are several physical processes to remove gas from galaxies 
in clusters, with subsequent starvation and star formation quenching:
tidal interactions between galaxies,  or
tidal stripping from the cluster potential itself, interactions with
the hot intra-cluster medium (ICM) through ram pressure, turbulent or viscous stripping,
or also outflows from star formation of nuclear activity,
We review the observational evidence for all processes,  and
numerical simulations of galaxies in clusters which support
the respective mechanisms. This allows to
 compare their relative efficiencies, all along cluster formation.
\end{abstract}

\section{Introduction: mechanisms to remove matter from galaxies}

Among the various possibilities to 
explain the stripping of matter (gas and stars) from galaxies in clusters,
the principal actors can be classified in three groups:\\
\indent 1. Tidal forces: \\
\indent \indent  -- interaction with a companion, merger: in this case,
a correlation between morphological type (T) and density ($\Sigma$) 
should be expected, (T-$\Sigma$ relation) \\
\indent \indent  -- interaction with the cluster; then a correlation 
between type and radius in the cluster is expected (T-R relation)\\
\indent \indent  -- harassment due to numerous interactions at high velocity and density\\
\indent 2. ICM-ISM interactions: \\
ram pressure stripping, but also thermal evaporation,
    turbulent, viscous stripping; these are purely hydrodynamical 
mechanisms, and should affect only the diffuse gas. However, they
are acting simultaneously with the others, and relative roles are hard to
disentangle. Since they are efficient only when the cluster is formed, and the ICM
gathered,  have they enough time to act? Or have tides acted before? \\
\indent 3. Outflows due to violent events:\\
\indent \indent -- starbursts and winds\\
\indent \indent -- AGN jets and outflows\\

All these processes result in morphological
type changes for galaxies, and stripping of their gas,
therefore star formation quenching, or "starvation" as
is observed in clusters. The delicate issue is that many mechanisms
are able alone to account for the stripping/quenching, and very
specific tests have to be found to disentangle what is happening.

\section{Observational clues} 

\subsection{Intra-cluster diffuse light}

\begin{figure}[t]
\centerline{
\psfig{figure=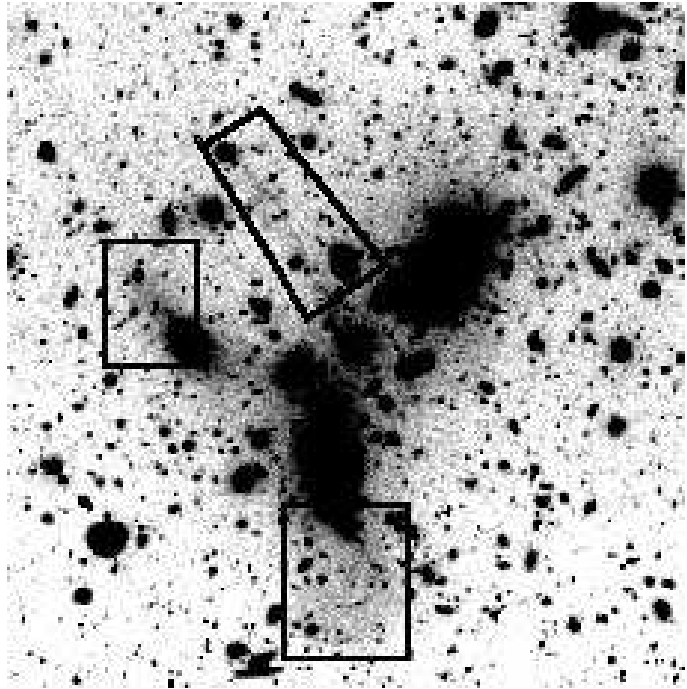,width=9cm,angle=0}
}
\caption{ Diffuse intra-cluster light is seen in this optical
 image of Abell 1914, from Feldmeier et al (2003).
Note the low-brightness common envelope around the central galaxies,
and the conspicuous tidal features.}
\label{fig1}
\end{figure}

One of the clear evidence of tidal interactions and stripping
is the existence of intra-cluster diffuse light (ICL): these 
intergalactic stars, stripped from their
parent galaxies by tidal interactions, represent
a large fraction of the total stellar mass of the cluster,
between 10-40\% (cf Figure 1, Feldmeier et al 2003).
Cluster images at low luminosity levels
show evidence of tidal debris in the form of plumes and 
arclike structures (example of the Centaurus cluster, 
Calc\'aneo-Rold\'an et al 2000).  The quantity  of ICL
does not appear to depend on cluster radius, but more
on the surface density of galaxies ($\Sigma$), which
favors the interactions between galaxies.

Although CCD images are now able to reveal ICL 
in most clusters, 
a large sensitivity for this diffuse component is gained
from Planetary Nebulae tracers, without the problems of
flat fielding, etc, since they are detected by emission lines
(Feldmeier et al 1998, Arnaboldi et al 2002) 
The intra-cluster stars have moderate metallicity (Durrell et al 2002),
which supports the scenario of their stripping from intermediate mass
galaxies.
These tidal debris and plumes are expected from simulations
of galaxy clusters (cf Dubinski 1998), even more prominent 
than what is observed. However, the background noise
dilutes the weaker features, explaining the difficulty
to observe them clearly (e.g. Mihos 2003).

\subsection{Larger fraction of blue galaxies at z=0.4 (BO effect)}

It has been known for a long time that  
there exists in clusters a larger fraction of blue galaxies at increasing 
redshift (Butcher \& Oemler 1978, 1984). These blue galaxies indicate 
much more star formation in the recent past, and correspond
to irregular shapes in the clusters. 
The existence in z=0.4 clusters of sign of tidal interaction/mergers
also confirm that clusters have evolved very recently: in the last
few Gyrs, there was a much larger fraction of perturbed galaxies,
late-types and starbursts, as if the cluster
had relaxed only since then.
Rings of star formation were much more frequent than 2-arms spirals,
contrary to what is found today (Oemler et al 1997).
These rings could be due to bars triggered in tidal interactions.
 Part of them could also be due to fast encounters, expected
in galaxy clusters, that lead to head-on collisions like the
Cartwheel. Alternatively galaxies, through harassment, could
be stripped at this epoch of their dark halos, de-stabilising disks.
and triggering more violent star formation. These tidal interactions
visible at z=0.4, must have profoundly and rapidly modified the galaxy 
morphologies, since at z=0.2, the evolution is almost terminated.
 Milder effects are observed by Balogh et al (1999) in an X-ray selected
sample of clusters (CNOC1), who suggest 
 a more gradual decline of star formation.

\subsection{Star formation rate versus environment and density}

In an H$\alpha$ line study of 11000 galaxies in the 2dF survey,
over 17 galaxy clusters, Lewis et al. (2002)  find the star formation
rate (SFR) increasing gradually from low values at the cluster centers, 
towards the field value at about 3 virial radii. They find a strong
correlation between SFR and local projected density, as soon as
the density is above 1 galaxy/Mpc$^2$, independent of the size
of the structure (i.e. also valid in groups).
G\'omez et al (2003) find also a strong SFR-$\Sigma$ relation
with the early data release of the SDSS, the SF-quenching effect
being even more noticeable for strongly star-forming galaxies. 
The same break of the SFR-$\Sigma$ relation is observed
at 1 galaxy/Mpc$^2$. This relation is somewhat linked to the morphological
type-density (T-$\Sigma$) relation, but cannot be reduced to it, since
at any given type, the SFR-$\Sigma$ relation is still observed. 
This strong relation valid even outside cluster cores is 
a precious clue to derive the dominant mechanisms.

\subsection{Morphological segregation}

\begin{figure}[t]
\centerline{
\psfig{figure=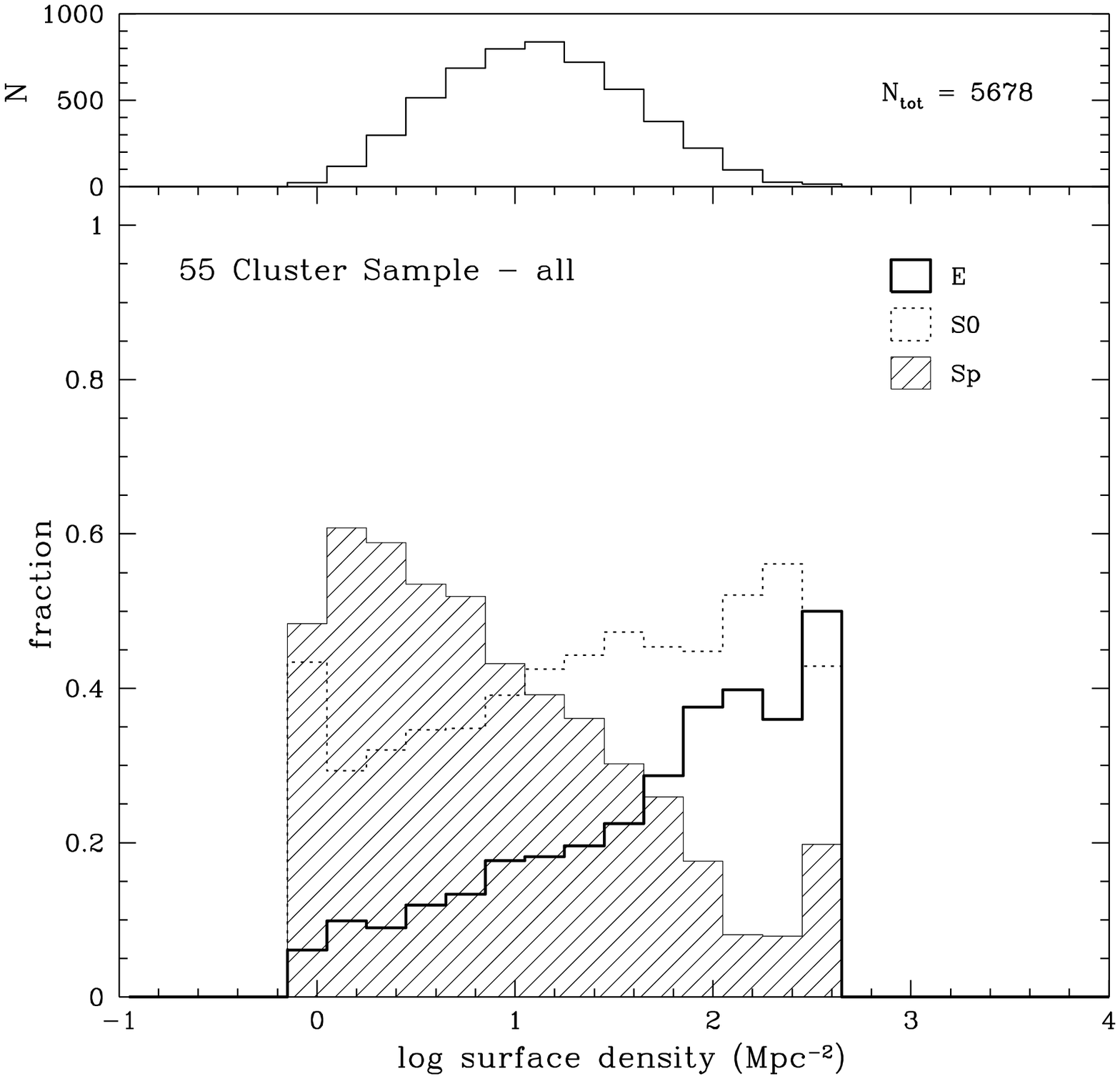,width=6.5cm,angle=0}
\psfig{figure=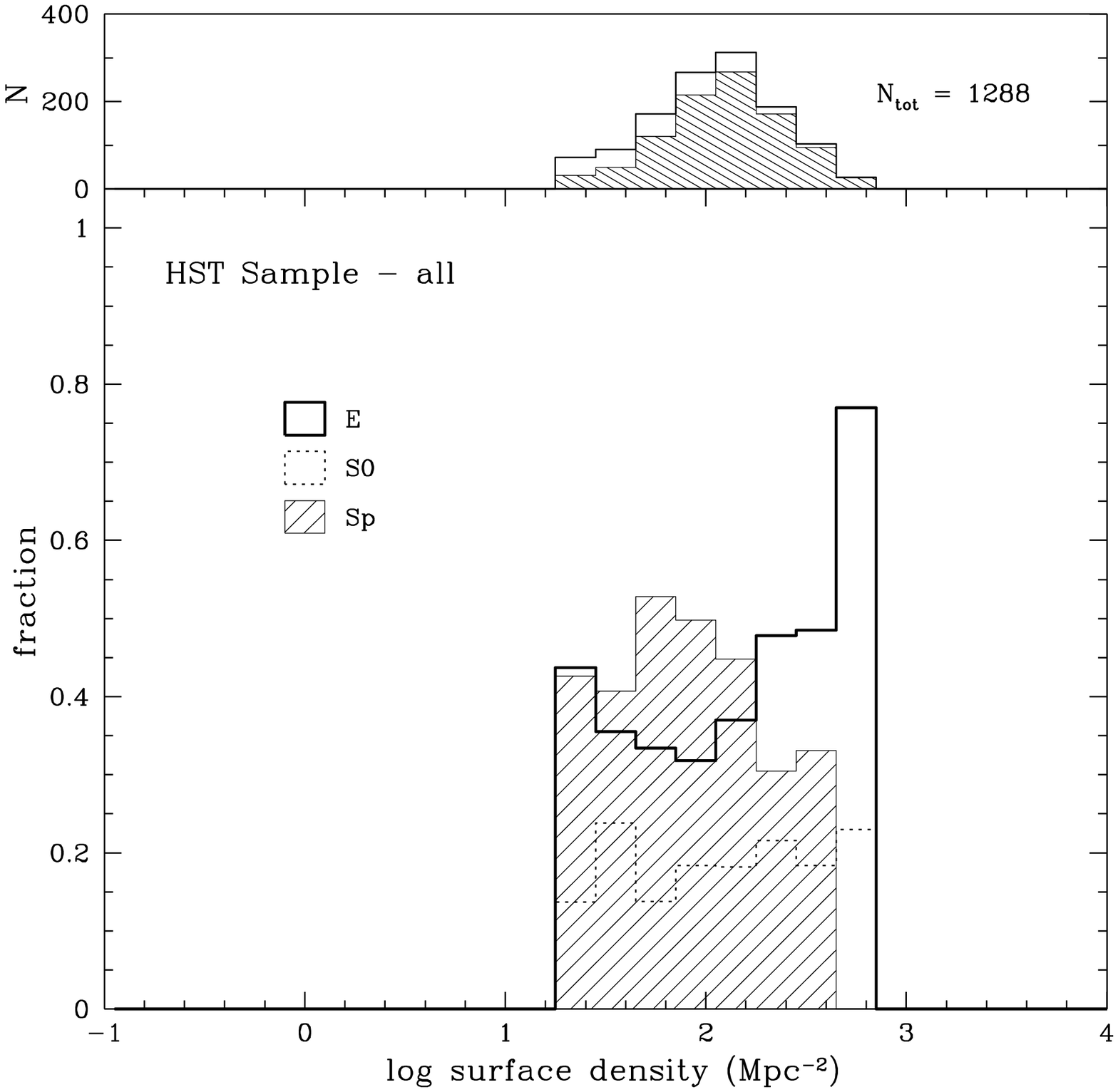,width=6.5cm,angle=0}
}
\caption{ Evolution of the morphological segregation
in clusters: the relation between type T and surface density $\Sigma$
for 55 clusters at z=0 from Dressler et al (1980, {\bf left}), 
and for 10 clusters at intermediate redshift 0.36 $<$ z $<$ 0.57,
from Dressler et al (1997, {\bf right}). 
 The histograms at top show the number density of
galaxies in each bin of $\Sigma$. Note the strong evolution
in the lenticular fraction (dotted histogram).}
\label{fig2}
\end{figure}

From the morphological segregation in nearby clusters,
drawn by Dressler et al (1980), it is now possible to 
see the evolution from about 5 Gyrs ago, at z=0.4
(Dressler et al 1997, Figure 2): at z=0, there was the
same T-$\Sigma$ correlation 
for relaxed or non-relaxed clusters, but it is 
no longer true at z=0.4.
 As main lines of evolution, there is at z=0.4 the
same fraction of ellipticals than at z=0, but a
much smaller fraction of S0s; at z=0.5, the fraction of lenticulars
is 3 times lower than now. This suggests that ellipticals
form early, before the cluster virialisation.
 In the hierarchical scenario, clusters form out of 
loose groups mergers, and it is likely that ellipticals are
the result of mergers in groups, before the formation of the cluster.

S0's are transformed from spirals in virialised clusters,
in a few Gyrs time-scale. The study of stellar populations,
and the spectral classes of cluster galaxies at z=0.4
reveals that star formation is quenched with respect
to the field (Poggianti et al 1999).  At z=0.4, passive and
post-starburst (E+A, or k+a) spirals are much more frequent than in the field.
It appears that the mechanism reponsible for that must
act on shorter time-scales than the mechanism reponsible
for the transformation into S0s.

\subsection{Density correlation}

\begin{figure}[t]
\centerline{
\psfig{figure=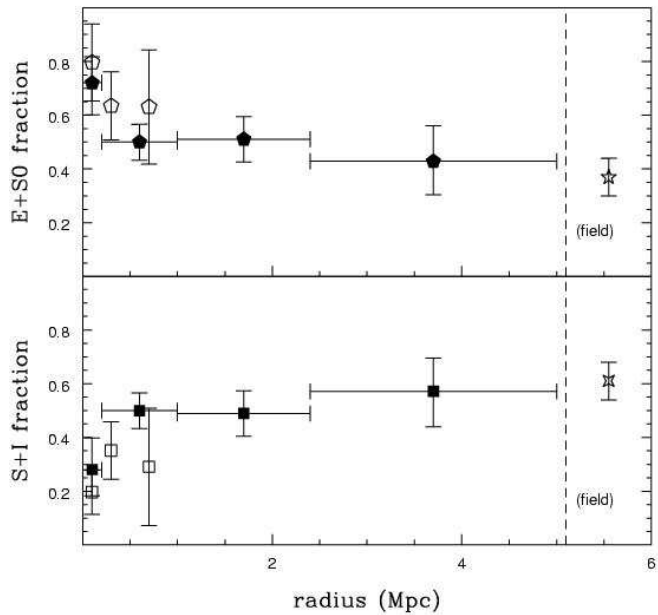,width=9cm,angle=0}
}
\caption{ The radial distribution of morphological types
(T-R relation)
for the cluster Cl0024+16, at z=0.4 ({\bf top:} early-types
E+SO, {\bf bottom:} spirals, from Treu et al. (2003).
The segregation goes up to 200kpc, well
inside the virial radius (which is 1.35 Mpc),
the field values are indicated at right.}
\label{fig3}
\end{figure}

In cluster regions where the density is not centrally symmetric,
it is possible to compare the 
morphology-radius (T-R) and morphology-density (T-$\Sigma$) 
relations. The latter (T-$\Sigma$) appears always better than
the T-R relation  (Treu et al. 2003, 
example of  Cl0024+16 at z=0.4). 
Galaxies are more aware of their local density than cluster location.

The morphological segregation as a function 
of radius is quite significant for radii lower than 200kpc (cf Figure 3).
The fraction of early-type galaxies drop steadily until 1 Mpc, or nearly
the virial radius. The correlation with radius is then weak, while
several over-dense regions have galaxies with morphology typical
of their high density. It seems that
gas depletion and morphological transformation are already well
advanced in groups, before forming the cluster.
The mild gradient in the morphological mix outside the virial radius 
could be due to harassment and starvation (ICM interactions
are not operating there). It is only upon arrival in the central
regions (R $<$ 200 kpc) that substructures are erased, as indicated by the
tight correlation between cluster radius and $\Sigma$. 

\subsection{HI deficiency}

Solanes et al. (2001) have recently made a data compilation on 1900 galaxies,
and conclude that about 2/3 of galaxy clusters are HI deficient in their centers.
Many observations demonstrate that the interaction with the hot gas (ICM-ISM interactions)
might be responsible for this HI stripping in cluster galaxies: large
deficiencies (up to a factor 100), deficiency
as a function of distance from the central X-ray peak, radial orbits of
the stripped galaxies, that allow them to explore the dense hot center.
In her review, van Gorkom (2003) describes convincing individual cases proving
ram pressure stripping, like galaxies in the Virgo cluster with perturbed and
reduced-size HI disks while the stellar disk is normal. Ram pressure stripping
appears quite efficient and rapid, playing a role in only one cluster crossing-time.

However, the correlation between HI deficiency and X-ray properties of the cluster
is not observed (L$_X$, T$_X$), which is surprising (Solanes et al 2001).
Instead, there are correlations with galaxy properties, 
early-type and probably dwarf spirals are more easily stripped than the
intermediate spiral types. That early-types are more stripped is not only
due to their position in the cluster, but their deficiency is larger
than from late spirals at each cluster radius, until 4 Mpc from the center.

\section{Simulation clues} 

\subsection{Tidal interactions and harassment}

Tidal interactions in galaxies were thought marginal because
of high velocities, and un-resonant interactions. However,
the large number of interations can accumulate perturbations
and truncation effects: this has been called 
``harassment'' by  Moore et al. (1996), i.e. frequent 
high-velocities close encounters.

Gnedin (2003) has recently simulated the formation
of galaxy clusters in the frame of a hierarchical cosmological
scenario, varying the cosmological parameters.
Tidal interactions determine the galaxy evolution, and are intensified by
the density irregularities, either the presence of massive galaxies,
or the infalling groups of galaxies, still not relaxed.
These substructures favor the interaction,  the typical frequency
being estimated at 10 interactions at 10kpc impact parameter 
per galaxy. Mergers occur essentially at the cluster formation,
 and are very rare today.  This means that elliptical galaxies
predate the cluster, as already found by Merritt (1984).
Later the tidal interactions can transform spirals to lenticulars, and 
explain their large fraction increase in the last Gyrs.
Tidal interactions truncate massive dark matter halos 
and thicken stellar disks, increasing disk stability and quenching
star formation. Dwarf galaxies can be totally disrupted.
 The collision rate per galaxy strongly decreases with time,
 from 8 to 2 per Gyr along the cluster life-time.

\subsection{Cluster tidal field}

The tidal field of the cluster itself has strong dynamical influence
on galaxies, in particular in their extended halos. First the dark matter
haloes are stripped, and form a common halo, but also the gas reservoirs that
replenish the interstellar medium of field galaxies all along their
lives, by accretion along gas filaments, is stripped also, and
this could easily explain the starvation, and gradual decline in
star formation of cluster galaxies. This phenomenon was first invoked
by Larson et al (1980), and simulated by Bekki et al (2001). 
The latter assume an accretion rate of 1 M$_\odot$/yr for a normal
field galaxy, and show that the tidal field of the cluster efficiently removes 
the gas reservoir from a galaxy, and consequently its fueling of star formation. 
This tidal truncation does not depend very strongly on the orbit of the
galaxy in the cluster, and the resulting SF-quenching is widespread 
through the cluster (contrary to what is expected from ICM interactions).
Once their gas reservoir is stripped, spiral galaxies will slowly be transformed
into lenticulars by harrasment, thickening and shortening
their stellar disk.

\subsection{Ram pressure and ICM-ISM interactions}

A large variety of models have been simulated, since the
phenomena associated with ICM interactions depend 
on many physical assumptions about the  small-scale
structure of the gas, instabilities, equivalent
viscosity, temperatures etc..
With an isothermal SPH gas model,
Abadi et al (1999) show that HI gas is effectively
stripped in the core of rich clusters,
for disks oriented perpendicular to the wind. Galaxies can lose
80\% of their gas, the final disk being restricted to 4kpc radius,
in a time-scale of 10$^7$ yrs. However, in the outer parts
of the cluster, or for inclined disks with respect to the wind, 
the process is much less efficient.

With a finite-difference code, Quilis et al. (2000)
show that viscous coupling could favour the stripping;
they show that a hole in the center of the galaxy (mimicking
the frequent HI depletion in the central regions), could
fragilize the gas disk, and enhance considerably the stripping
efficiency. Holes can have an influence at several scales.
If star formation has already formed shells through supernovae and winds,
ram pressure can enlarge the holes in the disk
(Bureau \& Carignan 2002).
Supernovae and winds alone are not efficient enough, except may be
in small dwarfs (Dekel \& Silk 1986, Martin 1998).

Vollmer et al (2001) use sticky particles
for the gas, and follow galaxies on their orbit through
the cluster: they show that the ICM has only a significant
action in the cluster core, and the stripped gas then
falls back on the galaxy, once its orbit gets out of the core.
Schulz \& Struck (2001) show that the stripping is a multi-step
process: the outer gas is quickly stripped, while
the inner gas is compressed, and forms a ring.
The compressed gas could give rise to triggered star formation
in a small starburst.

It is also possible that the gas reservoir required to replenish
the ISM of galaxies is hot and diffuse, as assumed by
Bekki et al (2002), who show through ram pressure simulations, 
that the halo will be efficiently stripped, if its density
is typically lower than 3 10$^{-5}$ cm$^{-3}$. In that case,
the global tidal stripping from the cluster and the ram pressure compete
to strip the gas reservoirs, and contribute to form passive or
anaemic spirals, that will slowly be transformed into S0s.

\section{Outflows: stellar and nuclear activity}

Enhanced stellar activity, when spiral galaxies infall into
the cluster, is observed (Kenney \& Yale 2002). If the 
galaxies have been stripped by the cluster global tide of
their halo, this can  favor the escape of the winds.
In a wide sample of galaxies, Kauffmann et al (2003)
have shown that star formation efficiency is
a strong function of surface density.  Low surface
density dwarf objects (LSB) are unevolved objects,
in which stellar feedback have prevented rapid
star formation, by ejection of their gas.
The energy of supernovae is enough to
disperse the gas, when the mass of the galaxy
has fallen below a threshold of 3 10$^{10}$ M$_\odot$,
the observed transition between low and high
surface density galaxies (Dekel \& Woo 2003).

Nuclear activity (AGN) could also be invoked to
provoke gas outflows, and remove gas from galaxies.
AGN have been found less frequent in cluster environment,
by a factor 5, with a frequency of only 1 percent,
compared to 5 percent in the field (Dressler et al. 1985). 
Recent observations show
however that they might be more frequent in X-rays, suggesting
an obscuration effect (Martini et al 2002).
The mechanical and heating energy of AGN has a strong 
feedback effect to reverse and self-regulate
the gas cooling in the centers of elliptical galaxies,  groups
and clusters  (e.g. Ciotti \& Ostriker 2001). AGN
feedback has been invoked to account for recent X-ray
observations incompatible with the old quasi-state cooling flow model:
absence of extremely cool gas ($<$ 1kev) in the center of cooling flow
clusters, presence of cold bubbles related to AGN radio lobes, etc...
 The amount of effectively cooling gas has been revised downwards.
The cooling flow could be intermittent, with alternate periods
of AGN activity, inflow and outflow.  A cold gas phase is
observed, as shown in the Abell 1795 cooling flow (figure 4).
This cluster center is not yet relaxed, with a cooling wake
triggered by the central cD motion (Fabian et al. 2001).
The X-ray data from the inner 200kpc indicate a mass deposition rate of 
about 100 M$_\odot$/yr (Ettori et al. 2002).

\begin{figure}[t]
\centerline{
\psfig{figure=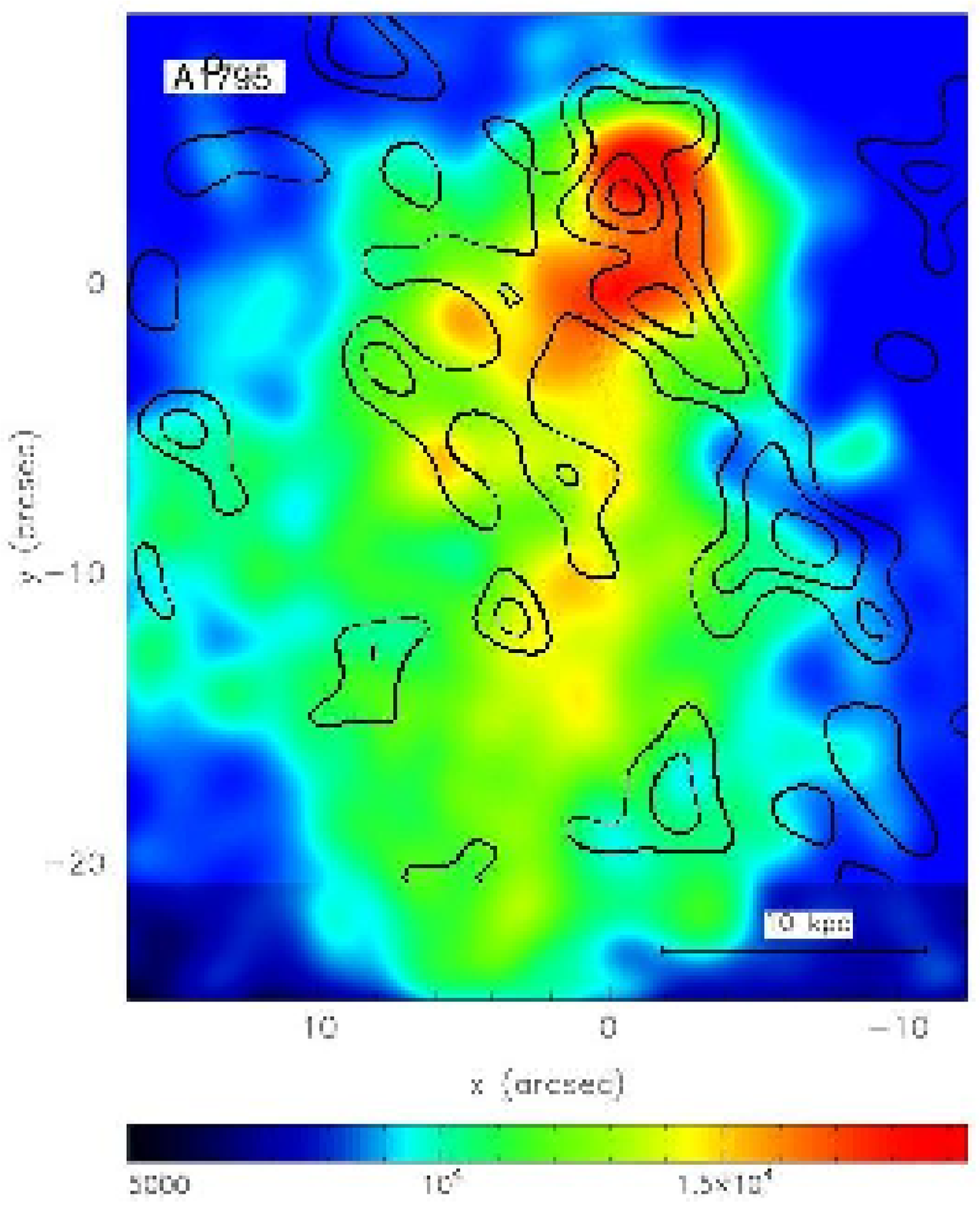,width=5cm,angle=0}
\psfig{figure=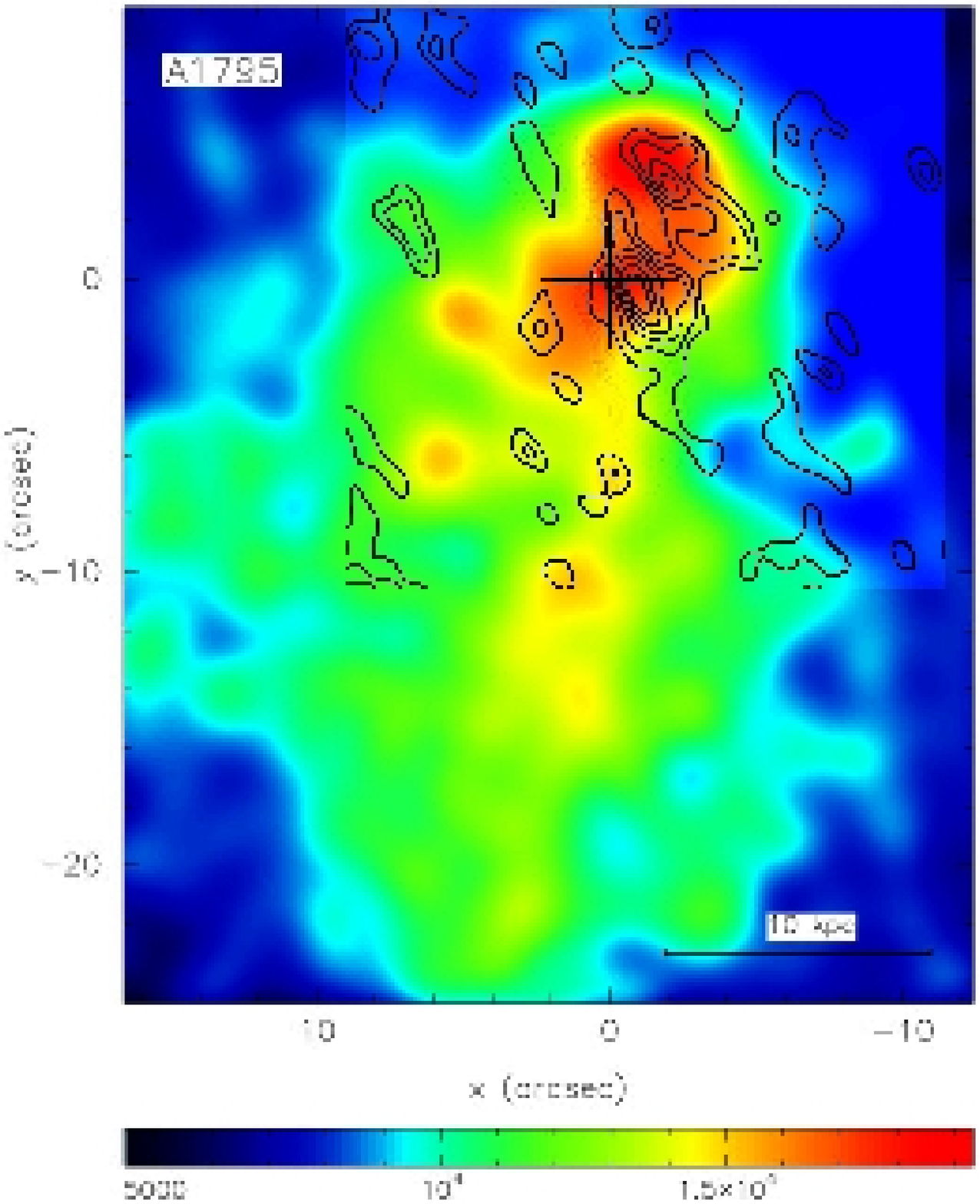,width=5cm,angle=0}
}
\caption{ Cold molecular gas associated to the cooling flow
of Abell 1795: contours of the CO(1-0) line ({\bf left}),
and CO(2-1) line ({\bf right}), superposed on the Chandra X-ray image,
by Fabian et al. (2001). The CO observations have been made
with the IRAM interferometer (Salom\'e \& Combes, 2003).}
\label{fig4}
\end{figure}

\section{Discussion}

\subsection{Chronology of events}

To find the different roles of the various stripping mechanisms
discussed above, it is essential to go back to the formation of the cluster,
and the chronology of events. The dense ICM required for efficient ram pressure 
is in place only after the formation of the cluster. Also the phenomenon
is proportional to the square of velocity dispersion, so virialisation 
should have occured, for all galaxies to acquire their high velocities.
In the hierarchical scenario of structure formation, groups form first,
with low velocity dispersion, and a large frequency of resonant tidal
interactions and frequent mergers. Giant ellipticals are formed at this
stage by spiral galaxy mergers in groups. This is part of the
morphological segregation, since the fraction of spirals decreases.

Progressively, the gas in galaxy haloes is stripped by tidal interactions
and heated by shocks to the virial temperature of the growing structure,
and the importance of the ICM interactions will grow. The ICM itself
is formed through gravitational forces.
A large fraction of the ICM gas comes from already processed galaxy disks,
since its metallicity is important (1/3 solar).

At the present epoch, when clusters are virialised and still relaxing,
the merger rate has fallen to zero (at least major mergers), and ellipticals
are only passively evolving. Ram pressure and the global cluster tide
are almost equally efficient to strip gas from infalling galaxies
(i.e. Bekki et al. 2002).

The fact that the T-$\Sigma$ relation is even tighter than the T-R relation,
combined to the extension of the T-$\Sigma$ relation from clusters down to
groups (Ramella et al. 1999, Helsdon \& Ponman, 2003) 
suggest that two-body gravitational interactions
are dominating. The transformation of spirals into S0s in the outer
parts of clusters, depending on the local density, does not pleade in 
favor of the ICM interactions as the main mechanism of the morphological
segregation.

Low density galaxies (dwarfs and LSB) entering a cluster can be
entirely disrupted by tidal shocks, they form streams of stars 
contributing to the intra-cluster light. 
Tidal phenomena lead similarly to a morphology-density relation
in loose groups, like that observed in the local group 
(Mayer et al 2001).

\subsection{Metallicity considerations}

In rich clusters, the mass of the hot gas can be much larger
than the mass of baryons in galaxies. Most of the baryons are
in the hot ICM, since it represents nearly the 
baryon fraction of the matter in the universe $f_b \sim 0.16$.
Given its relatively high metallicity, it is then
not surprising that most of the metals in a cluster comes from the 
ICM: there is about twice more Fe in the ICM than in galaxies
(Renzini 1997, 2003). 

Since the metals are synthetized in galaxies, this means that either
part of the hot gas comes directly from galaxies (by stripping,
or disruption), or that stellar winds and supernovae have enriched the
ICM. In fact, both sources of metals should be there, since metals expelled by
SNe would not be sufficient.

\begin{figure}[t]
\centerline{
\psfig{figure=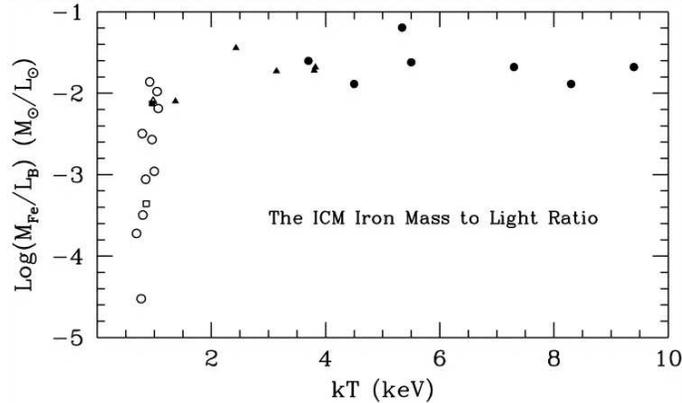,width=9cm,angle=0}
}
\caption{ The iron mass to light ratio of the ICM of clusters and groups as 
a function of their hot gas temperature, from 
Renzini (1997).}
\label{fig5}
\end{figure}

Figure 5 shows that
the iron mass-to-light ratio is about constant as a function of
the virial temperature of the structure (below 2 kev, the estimations
are less simple to derive, and there could be biases).
These observations suggest that metals are ejected via winds
(SNe or AGN), not ram pressure, since there is no dependence on 
richness, or cluster velocity dispersion,
but only the dispersion of individual galaxies (Renzini 2003).
There is the same M$_{Fe}$/L$_B$ in clusters and 
galaxies, implying the same processing in 
clusters than in the field.

Ellipticals in the field or in clusters have the same properties,
confirming that these galaxies have been formed before
the clusters. Since the stellar mass is essentially in
elliptical galaxies today (3/4 of the mass in spheroids, 
1/4 in disks, less than  1\% in Irr, Fukugita et al. 1998,
and Es are dominating even more in clusters), 
most cluster stars are formed before cluster formation.

Stars and corresponding metals have likely been made at z$\sim$ 2-3,
at the peak of stellar activity in the universe.
Part of the iron has been made through SNII quickly, 
then by SNIa, 1Gyr after the main episodes of star formation.
Clusters have not lost iron, nor accreted pristine material
since the ratio between $\alpha$ elements (made in SNII) and iron is 
about solar in the ISM of all clusters (with no or little
variation from cluster to cluster). This means that the metals come
from normal stellar nucleosynthesis, with similar ratio
of type Ia to type II SNe, as well as the same global IMF, etc..

Also, at the present time SNIa continue to enrich in Fe the medium
essentially near the cD at the center; it has been
shown that clusters have almost negligible metallicity
gradients, except those with cooling flows,
and bright central galaxies (B\"ohringer, this symposium).

\section{Conclusion}

The various clues brought by observations and simulations
allow to draw the following conclusions, about
the stripping mechanisms and the recycling of matter in
clusters:\\
 -- Tidal forces are at play from the beginning of
structure formation, and are the dominant factor
in the T-$\Sigma$ relation; they need long time-scales (1Gyr) in
groups that will finally (recently, z$\sim$1) merge in a cluster\\
--  ICM interactions have entered in action only recently 
(after the virialisation of rich clusters); they are very efficient
in HI stripping from galaxy disks infalling today into
the cluster in nearly radial orbits. The corresponding
stripping time-scales are very short (10$^7$-10$^8$ yr).\\
-- Starburst and AGN winds have played a role 
continuously (with a peak at z=2-3), and especially
in the metallicity distribution\\
-- E's have formed in groups before the cluster, 
S0's have been transformed from spirals in rich clusters 
in the last 5 Gyr;  their gas reservoir has been stripped
through both the global tide of the cluster and the ICM
interaction, and their star formation quenched. Harassment
progressively truncates and thickens their disks.

The difficulty to derive exactly the cluster evolution
is that the mechanisms may act simultaneously, and also
they show some duality: for instance,  
on one hand tidal interactions trigger some star formation,
leading to the BO-effect, and observed larger fraction
of blue galaxies at z=0.4 in clusters, but also the tidal
truncation of gas reservoirs implies a gradual decline in
star formation in the last Gyrs. 


\end{document}